\documentclass[reprint,twocolumn]{revtex4-1}
\usepackage{graphicx}
\usepackage[OT1]{fontenc}
\usepackage[utf8]{inputenc}
\usepackage{bm,amsmath,amssymb,amsbsy,amstext}
\usepackage[colorlinks=false]{hyperref}
\usepackage{color}

\begin{document}

\title{
Integration of external electric fields in molecular dynamics simulation models for resistive switching devices}

\author{T. Gergs}
\affiliation{Institute of Theoretical Electrical Engineering, Ruhr University Bochum, Bochum, 44787, Germany}
\author{S. Dirkmann}
\author{T. Mussenbrock}
\affiliation{Electrodynamics and Physical Electronics Group, Brandenburg University of Technology Cottbus-Senftenberg, Cottbus, 03046, Germany
}

\date{\today}

\begin{abstract}

Resistive switching devices emerged a huge amount of interest as promising candidates for non-volatile memories as well as artificial synapses due to their memristive behavior. The main physical and chemical phenomena which define their functionality are driven by externally applied voltages, and the resulting electric fields. Although molecular dynamics simulations are widely used in order to describe the dynamics on the corresponding atomic length and time scales, there is a lack of models which allow for the actual driving force of the dynamics, i.e. externally applied electric fields. {This is due to the restriction of currently applied models to either solely conductive, non-reactive or insulating materials, with thicknesses in the order of the potential cutoff radius, i.e., 10 \r{A}. In this work, we propose a generic model, which can be applied in particular to describe the resistive switching phenomena of metal-insulator-metal systems.} It has been shown that the calculated electric field and force distribution in case of the chosen example system Cu/a-SiO$_2$/Cu are in agreement with fundamental field theoretical expectations. 
\end{abstract}


\maketitle

\section{INTRODUCTION}
\begin{figure}[t!]
\begin{centering}
\includegraphics[width=8.5cm]{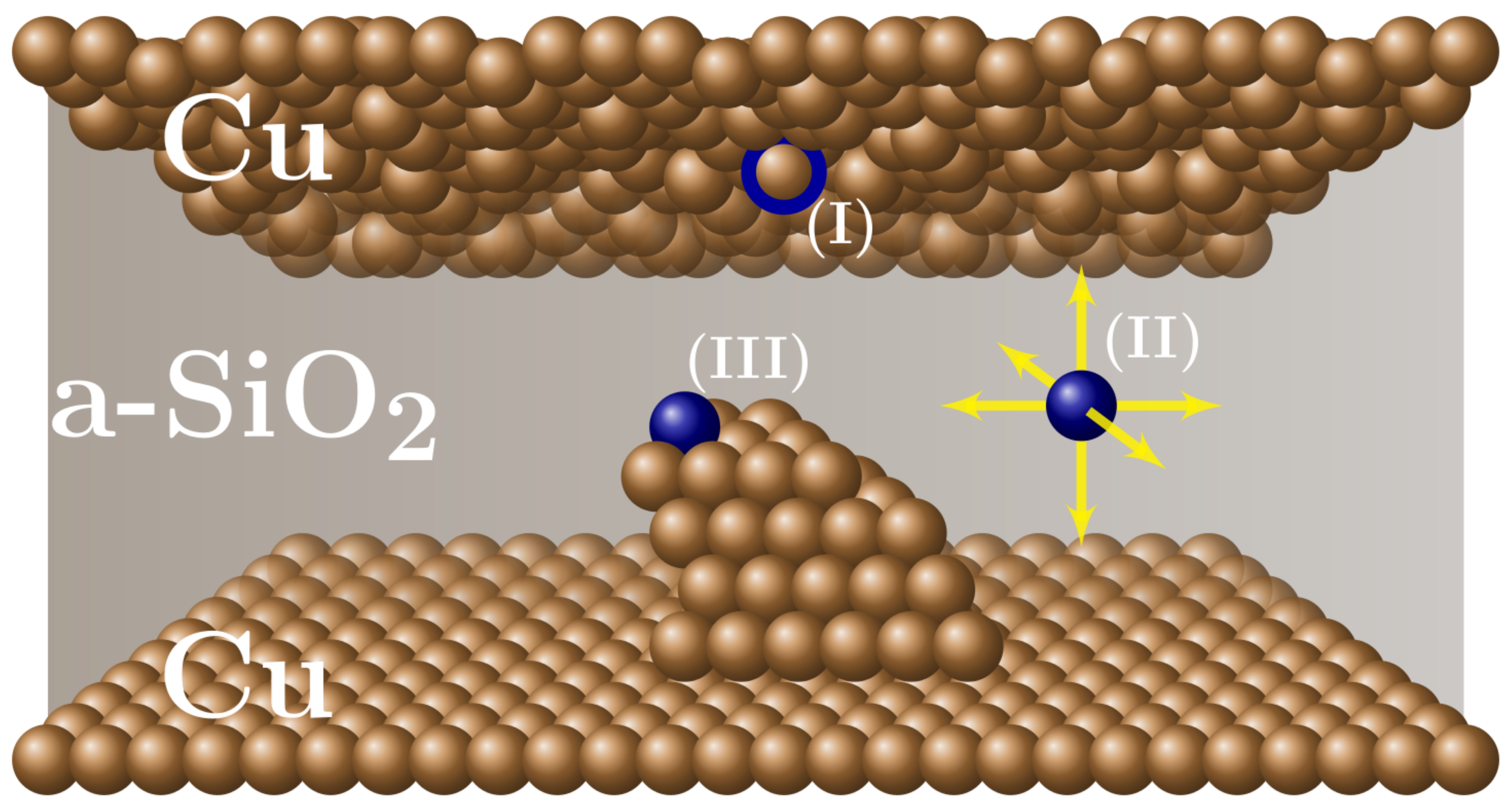}
\end{centering}
\caption{
Cu/a-SiO$_{2}$/Cu electrochemical metalization cell with the fundamental processes of oxidation (I), drift (II),  and reduction of Cu-atoms/-ions (III).
\label{fig: SchemaZeichnung} 
}
\end{figure}

Physical neural networks are typically investigated to utilize the efficiency of the biological information processing. The function of the neural synapses are emulated by resistive switching devices. Of paramount importance in this and in the context of non-volatile memories is their memristive behavior \cite{Waser2009, Yang2013}. Many of these two terminal devices have a conceptually simple metal-insulator-metal structure and rely on ionic mechanisms on the nanometer length scale, one of which is electrochemical metallization indicated in Fig \ref{fig: SchemaZeichnung}: (I) oxidation of metal atoms at the active electrode, (II) drift of the resultant ions within the dielectric and (III) reduction at the opposite electrode. Resistive switching is a consequence of the formation and re-formation of conductive metal filaments inside the insulator. It has been shown, that the processes on the device length scale and on the experimental time scale, i.e. microseconds to minutes can be successfully simulated by means of kinetic Monte Carlo (KMC) methods \cite{Menzel2015understanding, dirkmann2015kinetic,Dirkmann2017}. However, KMC simulations rely inherently on the identification of all important physical and chemical processes and their corresponding transition rates. These informations are crucial for reliably modelling the device dynamics and can be gathered to a great extent by classical molecular dynamics (CMD) simulations.

Within CMD approaches Newton's equations of motion are solved subject to interaction potentials, which resemble the respective quantum mechanical behavior. Therefore, the CMD method provides the basis for a deeper understanding of the physical processes on atomic time scales \cite{onofrio2015atomic}. It is however important to note that the dynamics of resistive switching devices and their functionality are mainly driven by externally applied voltages and therefore externally applied electric fields. {Up to now, only a small number of CMD models which allow for externally applied electric fields and reactive systems have been reported \cite{dapp2013redox, nistor2009dielectric, assowe2012reactive, onofrio2015atomic}.} 

{However, these models are not suitable for the simulation of reactive metal-insulator-metal systems, in which the interfacial forces are of major importance. Nistor et al. proposed a polarization model for the split charge equilibration (SQE) formalism which was shown to be suitable for simulations of capacitor geometries of a single dielectric material (without metal electrodes) under the influence of a constant external electric field \cite{nistor2009dielectric}. Though for the simulation of resistive switching devices the electric field at the neglected metal/insulator interface is of particular interest. Dapp et al. combined the SQE method with the model of an external circuit for a simplified simulation of a battery where an electrolyte is placed between two metal electrodes \cite{dapp2013redox}. The combination of these two models could in principle be used for resistive switching devices. However, the long range Coulomb interactions used for these example cases are not compatible with reactive potentials where interactions among atoms are limited by cutoff radii. These potentials are however required for the simulation of the respective phenomena at the ternary Cu/a-SiO$_2$ interface. Assowe et al. used the reactive force field (ReaxFF) in combination with a modification of the charge equilibration (QEq) method to include polarization for the investigation of the Ni(111)/H$_2$O interface under the influence of a constant external electric field \cite{assowe2012reactive}. The source of the external electric field, which would be of major importance for resistive switching, is however neglected. In addition the respective modification of the QEq method resembles global charge transfer polarization currents. This means that the system's response to an external electric field resembles always the response of an ideal conductor. This is very appropriate for the respective investigation of the Cu/H$_2$O interface. However, this implicates problems for the present Cu/a-SiO$_2$/Cu material system. Here the application of the QEq method would allow for a current of free charges from one electrode to the other. At the same time the polarization of the insulator a-SiO$_2$ would be solely the polarization of a metal. This would lead to very high electric fields at the interfaces and a vanishing electric field inside the insulator. Therefore the oxidation would be over- and the drift inside the insulator underestimated. The fluctuating charge model QTPIE (charge transfer polarization current equalization), which can be described as an extension of the QEq method, can be used as model for the appropriate polarization of insulators \cite{chen2007qtpie, chen2008unified, chen2009theory}. Therefore one may implement the external voltage as combination of the QEq (for the metal electrodes) and the QTPIE method (for the insulator), both modified to include polarization. This might in fact provide the correct charge distribution. However due to the interaction limitations of reactive bond order potentials the insulator thickness would have to be smaller than the respective cutoff radius. This would allow in the end for insulator thicknesses up to 10 \r{A}. However, for electrochemical metallization devices insulator thicknesses up to 100 \r{A} are typically used. 
}

In order to overcome these limitations and remedy the lack of a generic model, we propose an alternative scheme for electric fields in self-consistent CMD simulations of metal-insulator-metal systems. The model is based on the separated treatment of the different kinds of electric forces in field driven metal-insulator-metal systems. Thereby, the fluctuating charge model QTPIE has been chosen to describe the polarization of the insulator. Overall, the proposed model is compatible with any charge equilibration based CMD potential and the widely used molecular dynamics simulation tool LAMMPS \cite{plimpton1995fast}.

The model has been validated by using the reactive force field ReaxFF to simulate a Cu/a-SiO$_2$/Cu material stack as an example system. We have chosen this particular system as Cu/a-SiO$_2$ is an often used electrode/insulator material composition for electrochemical metallization devices \cite{Schindler2009Electrode}. {ReaxFF itself is based on bond-order dependent interactions, which are used to describe harmonic bonds. The potential energy of the system is mostly described by two-, three- and four-body terms, as well as van der Waals and modified Coulomb interactions \cite{van2001reaxff}. Furthermore, lone pairs as well as under- and over-coordinated atoms are punished by means of an additional potential energy. Here, the fluctuating and environment dependent charge distribution is calculated self-consistently with the QTPIE method, which itself depends mostly on the atoms' electronegativity as well as hardness. The extensive parameter sets for the ReaxFF potential and point charge model are typically fitted to experimental findings and ab initio calculations to suit the potential for its particular application.}
Ultimately, it has been shown, that the proposed integration for the electric field is particularly applicable to investigate resistive switching in electrochemical metallization cells due to the dependency of the respective fundamental processes on the electric field. Due its generic character, the model can be easily utilized for any field driven metal-conductor-metal or metal-insulator-metal system.

\section{Electric field and force model \label{sec:Implementation}}

In metal-insulator-metal resistive switching devices the most important physical and chemical processes are driven by external electric fields. The related electric forces which act on interfacial metal atoms can be split into three parts, which will be discussed below. For this ansatz the electric field at the position of the $i$-th atom is decomposed into two parts, i.e., $\bm{E}_{\mathrm{ext},i}$ and $\bm{E}_{\mathrm{int},i}$. The latter is given as $\bm{E}_{\mathrm{int},i} = \boldsymbol{E}{}_{\textrm{dep},i} + \bm{E}_{0,i}$. $\bm{E}_{0,i}$ describes Coulomb interactions without any externally applied voltage and $\boldsymbol{E}{}_{\textrm{dep},i}$ describes the depolarization field.

First, the forces acting on the free charges at the electrodes as a result of the applied voltage are expressed by means of the Maxwell stress tensor $\boldsymbol{T}(\boldsymbol{E}_{\textrm{int}},\boldsymbol{E}_{\textrm{ext}})_{i}$ and $\boldsymbol{T}(\boldsymbol{E}_{\textrm{ext}})_{i}$ as a function of the respective electric fields
\begin{align}
\boldsymbol{F}_{\mathrm{free,}i} &=  \int_{V}\rho_\mathrm{free}\boldsymbol{E}_{i}d^{3}r,\\
&=\oint_{\partial V}\epsilon_{0}\left[\boldsymbol{T}(\boldsymbol{E}_{\textrm{ext}})_{i}+\boldsymbol{T}(\boldsymbol{E}_{\textrm{int}},\boldsymbol{E}_{\textrm{ext}})_{i}\right]\cdot\boldsymbol{n}da\label{eq:Fel Herleitung 5.0}.
\end{align}
{These are charges which are induced at the electrodes solely due to the external voltage, neglecting the dielectric and thus generating the external electric field $\bm{E}_{\mathrm{ext}}$. Consequently, this term is zero for any metallic ion which is not bounded to an electrode and any non-metallic atom inside the insulator. Second, the electric forces due to the external electric field acting on the polarization charge distribution at the electrodes and in the insulator are described by $q_{\textrm{pol,}i}\boldsymbol{E}_{\textrm{ext,}i}$.} The polarized charge of the $i$-th atom is given here by $q_{\mathrm{pol,}i}$. Third, the Coulomb forces on these charges are described by $q_{\textrm{pol,}i}\boldsymbol{E}_{\textrm{int,}i}$. Altogether the electric forces on interfacial metal atoms can be expressed as 
\begin{align}
\boldsymbol{F}_{i} &= \oint_{\partial V}\epsilon_{0}\left[\boldsymbol{T}(\boldsymbol{E}_{\textrm{ext}})_{i}+\boldsymbol{T}(\boldsymbol{E}_{\textrm{int}},\boldsymbol{E}_{\textrm{ext}})_{i}\right]\cdot\boldsymbol{n}da\dots\nonumber\\&\qquad\qquad\dots+q_{\textrm{pol,}i}\boldsymbol{E}_{\textrm{ext,}i}+q_{\textrm{pol,}i}\boldsymbol{E}_{\textrm{int,}i}\label{eq:Fel Herleitung 5}.
\end{align}

{The subsequent calculation of the electric forces acting on the $i$-th atom at the electrodes or in the insulator can be divided into three complementary parts:} i) The calculation of the external electric potential $\Phi_{\mathrm{ext}}$ and field $\bm{E}_{\mathrm{ext}}$, ii) the calculation of the polarized charges $q_{\mathrm{pol,}}$ and the internal electric field $\bm{E}_{\mathrm{int}}$, and iii) the calculation of the Maxwell stress tensor $\boldsymbol{T}(\boldsymbol{E}_{\textrm{int}},\boldsymbol{E}_{\textrm{ext}})$ and $\boldsymbol{T}(\boldsymbol{E}_{\textrm{ext}})$. In the following sections, these three parts are separately discussed.

\subsection{Calculation of the external electric field\label{sub:External-electric-field}}

The calculation of the potential $\Phi_{\mathrm{ext}}$ due to the externally applied voltage is based on solving the Laplace equation,
\begin{gather}
\boldsymbol{\nabla}^{2}\Phi_{\textrm{ext}}=0,
\label{Laplace}
\end{gather}
subject to appropriate boundary conditions. The application of Dirichlet boundary conditions requires a continuous representation. {First, the metal atoms at the interfaces are detected self consistently by a combination of cluster and coordination number analysis. Second, the $z$ coordinate of these atoms are interpolated on a grid in the $x-y$ plane. Then, the electric field can be calculated on coarser grid points solving}
\begin{gather}
\boldsymbol{E}_{\textrm{ext}}=-\boldsymbol{\nabla}\Phi_{\textrm{ext}}.
\label{Field}
\end{gather}
This is appropriate since the electrostatic approximation of Maxwell's equation is justified.

\subsection{Calculation of the polarized charges and of the internal electric field \label{sub:Polarization}}

The externally applied electric field influences the overall atomic charge distribution. Therefore, a model for the fluctuating charges is needed. {In the following, suitable fluctuating charge models, their similarities as well as differences, are described briefly. Further details can be found in the respective literature. The appropriate application of these models is crucial for a sophisticated implementation of electric fields.}

In the widely used charge equilibration (QEq) method, the energy of the system is expanded in the charge distribution and truncated after the second order. The first- and second-order terms are identified with the electronegativity and the hardness, respectively. The equalization of the electronegativity or the chemical potential by charge transfer within the bonded system leads to the charge distribution, which minimizes the electrostatic energy under the constraint of charge neutrality. The corresponding system of linear equations can be solved by applying the method of Lagrange multiplier \cite{rappe1991charge, nakano1997parallel}. This formalism implies global charge transfer currents between the atoms, where distances represent no limitations. Nevertheless, this method provides in general a satisfactory description for many cases without any external perturbation. The response to an external electric field is though always a global rearrangement of the atomic charges, comparable to an ideal conductor. As a consequence with this method the polarization of insulators is only modeled insufficiently and the net electric field inside the insulator is zero \cite{nistor2009dielectric}. 

Since in the present case the electrodes are typically separated by a thin insulator, an effective and efficient model for fluctuating charges is needed. Therefore, the QEq method is extended to the charge transfer polarization current equalization (QTPIE) method, where polarization currents between atoms are locally limited \cite{chen2007qtpie}. The formalism itself is similar to the formalism of the QEq method, whereas the atomic charge variables $q_i$ are initially substituted with the corresponding sum of charge transfer variables $p_{ij}$, which describe the charge transfer from atom $j$ to atom $i$. The application of charge transfer variables is meanwhile similar to the SQE approach \cite{nistor2006generalization}. However, in this work an additional term for long range charge transfer currents between the atoms has been included. The subsequent expression for the electrostatic energy as a function of the transfer charge variables $p_{ij}$ can be reformulated by means of charge conservation as a function of the atomic charge variables, $q_{i}=\sum_{j}p_{ji}$ \cite{chen2008unified}. The resultant equation for the electrostatic energy is then given by
\begin{multline}
W_{\textrm{es}}=\sum_{i}\chi_{\textrm{eff,}i}q_{i}+\frac{1}{2}\sum_{i}J_{i}q_{i}^{2}\dots \\+\sum_{ij}\mathsf{\frac{1}{4\pi\epsilon_{0}}}\textrm{Tap}(r_{ij})\frac{q_{i}q_{j}}{(r_{ij}^{3}+\gamma_{ij}^{-3})^{-3}}.\label{eq:QEq Energie-ReaxFF}
\end{multline}
This equation resembles the initial expansion of the energy in the charge distribution, whereas for simplicity the charge independent term is omitted. {The hardness is described by $J_i$.} The tapering function $\textrm{Tap}(r_{ij})$ is used to avoid cutoff radius generated discontinuities within the potential energy surface. In addition the $\gamma_{ij}$ value is introduced in order to allow for screening effects. The first term corresponds with the first-order of the beforehand mentioned expansion and is responsible for the spatial limitations of the charge transfer between atoms by the definition of the effective electronegativity
\begin{gather}
\chi_{\textrm{eff,}i,0}=\frac{\sum_{j}(\chi_{i}-\chi_{j})S_{ij}}{\sum_{j'}S_{ij'}}. \label{eq:QTPIE Chi Eff}
\end{gather}
 This atomic property is computed by weighting the differences of an atom and its surrounding atoms' electronegativities with the corresponding overlap integral of Gaussian-type orbitals (GTOs) $S_{ij}$. The latter has been chosen in favour of Slater-type orbitals due to its simpler computation \cite{helgaker2000molecular}.

In case of electric fields, the change in the electrostatic energy is taken into account by the following modification of the effective electronegativity
\begin{gather}
\chi_{\textrm{eff,}i}=\frac{\sum_{j}(\chi_{i}-\chi_{j}+\Phi_{j})S_{ij}}{\sum_{j'}S_{ij'}}-\Phi_{i}.\label{eq:chieff2}
\end{gather}
This is a slightly modified version compared with the model presented by Chen et al. The detailed derivation of the energy and the effective electronegativies can be found in \cite{chen2009theory}. Since this ansatz leads on one hand to the polarization inside the dielectric and on the other hand to a corresponding compensation at the electrodes, this kind of polarization model is well suited for the investigation of voltage driven memristive devices.

The second and the third term of \eqref{eq:QEq Energie-ReaxFF} describe the second-order corrections of the energy expansion, whereas the Coulomb self-energy and Coulomb interactions are treated separately. The equation for the latter originates from the electron equilibration method and is also used for the ReaxFF potential \cite{mortier1986electronegativity, janssens1995comparison}. Subsequently the electric field can be directly calculated from the latter contribution to the electrostatic energy,
{
\begin{gather}
\boldsymbol{E}_{\textrm{int},j}=-\sum_{i\ne j}\mathsf{\frac{1}{4\pi\epsilon_{0}}}\bm{\nabla}\left(\textrm{Tap}(r_{ij})\frac{q_{i}}{({r}_{ij}^{3}+\gamma_{ij}^{-3})^{-3}}\right).\label{eq:Einteq}
\end{gather}
}

\subsection{ Calculation of the Maxwell stress tensor \label{sub:Sources}}
If both the external and internal electric fields are known, the components of the Maxwell stress tensors can be calculated by
\begin{eqnarray}
&T_{mn}&(\boldsymbol{E}_{\textrm{ext}}) = E_{\textrm{ext},m}E_{\textrm{ext},n}-\frac{E_{\textrm{ext},m}E_{\textrm{ext},m}}{2}\delta_{mn},\\
&T_{mn}&(\boldsymbol{E}_{\textrm{int}},\boldsymbol{E}_{\textrm{ext}})  =  E_{\textrm{int},m}E_{\textrm{ext},n}\nonumber\dots \\
&\textcolor{white}{T_{mn}}&\textcolor{white}{(\boldsymbol{E}_{\textrm{ext}}) = E_{\textrm{ext},m}E_{\textrm{ext},n}}
-\frac{E_{\textrm{int},m}E_{\textrm{ext},m}}{2}\delta_{mn},
\end{eqnarray}
{using Einstein's summation convention with the Kronecker symbol $\delta_{mn}$. Subsequently, the bounding surface integral is evaluated at the beforehand calculated Cu/a-SiO$_2$ interfaces which were previously used for the application of the Dirichlet Boundary condition. The forces on the surface elements are then distributed onto the respectively nearest metal atoms.}

\section{Validation of the model \label{sec:RESULTS}}

\begin{figure}[t!]
\begin{centering}
\includegraphics[width=8cm]{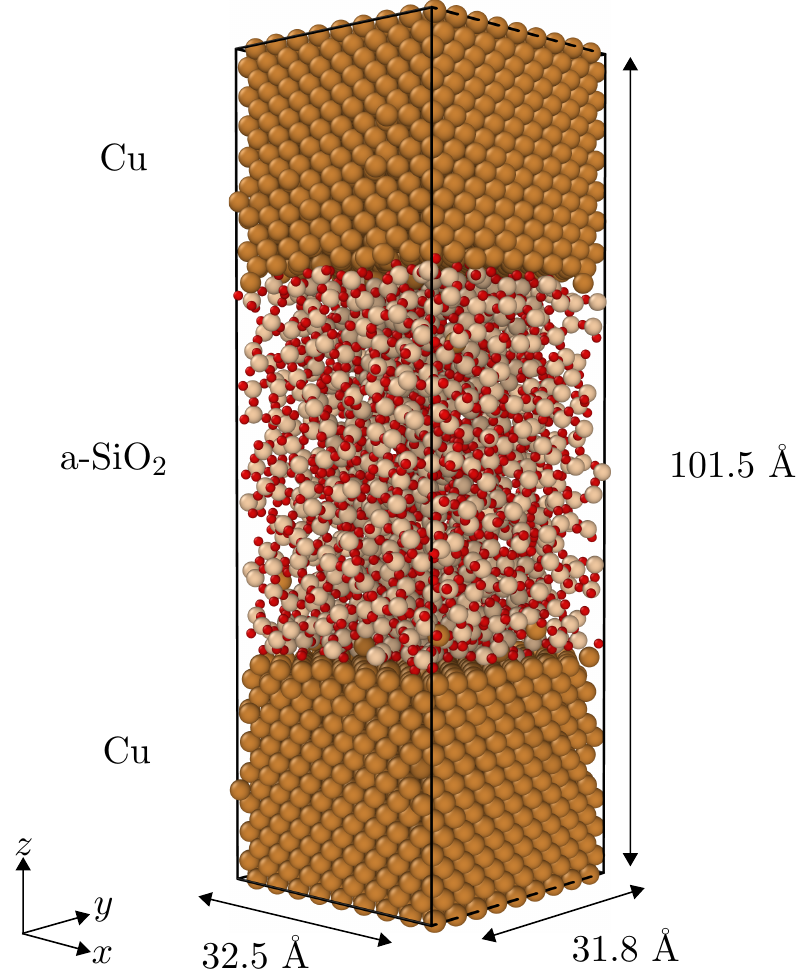}
\end{centering}
\caption{
Cu/a-SiO$_{2}$/Cu material system with periodic boundaries along the $x$ and $y$ direction. \label{fig: Struktur}}
\end{figure}

To validate the developed model, the electric field and force distribution is calculated within an exemplary Cu/a-SiO$_2$/Cu system. Dirichlet boundary conditions have been applied in $z$ direction, whereas in $x$ and $y$ direction periodic boundary conditions have been chosen (see Fig. \ref{fig: Struktur}). For all performed simulations a suitable potential has been used within the chosen molecular dynamics simulation code LAMMPS. The actual field and force calculation model is consistently coupled with LAMMPS via a Python interface \cite{van2010development, van2003reaxffsio, onofrio2015atomic, plimpton1995fast}. At first, the preparation of the example system is described. Afterwards, a suitable diagnostic method to measure electric fields is developed. Finally, the validation results for our model are presented and discussed.

\subsection{System preparation\label{sub:System-preparation}}

The a-SiO$_2$ layer has been prepared according to the procedure published by Fogarty et al. \cite{fogarty2010reactive}. For this, the implemented versions of the Nos\'e-Hoover thermostat, the Berendsen thermostat and the Berendsen barostat have been used \cite{shinoda2004rapid, tuckerman2006liouville, martyna1994constant, parrinello1981polymorphic, berendsen1984molecular}. Afterwards, the a-SiO$_2$ has been replicated in $z$ direction and placed between two crystalline copper blocks with a distance of 1.5 \r{A}. In the next step, the atomic positions were iteratively adjusted to minimize the overall energy of the system. Since a further annealing of the whole system due to an application of a single Nos\'e-Hoover thermostat leads to a temperature drift between the metal blocks and the insulator, each material layer of the trilayer system was coupled to a separate Nos\'e-Hoover thermostat. The atomic affiliation to the layers was determined dynamically by a cluster-analysis with a cutoff radius of 3 \r{A}. After annealing the system at 700 K for 75 ps the three Nos\'e-Hoover thermostats were replaced by three Berendsen thermostats to quench the temperature to 300 K with a rate of 25 K/ps. Finally, the three Berendsen thermostats were exchanged with three Nos\'e-Hoover thermostats and the system was simulated for 50 ps to reach a steady state.

\subsection{Electric field diagnostics\label{sub:Measurement-methods-for}}

The diagnostics for the electric field inside the system is based on the assumption that the force on an arbitrary atom inside the dielectric can be expressed as
\begin{align}
&\boldsymbol{F}_{i}|_{\chi_{\mathrm{eff},i}}  =  q_{\textrm{pol,}i}(\boldsymbol{E}{}_{\textrm{dep},i}+\boldsymbol{E}_{\textrm{ext,}i})+\boldsymbol{F}_{0,i},\label{eq:Diagnostic pos Q}\\
&\boldsymbol{F}_{i}|_{-\chi_{\mathrm{eff},i}}  =  q_{\textrm{pol,}i}(\boldsymbol{E}{}_{\textrm{dep},i}-\boldsymbol{E}_{\textrm{ext,}i})+\boldsymbol{F}_{0,i}\label{eq:Diagnostic neg Q}.
\end{align}
The term $\bm{F}_{0,i}$ is the force on an atom without any applied voltage. \eqref{eq:Diagnostic neg Q} is implemented by the inversion of the sign of the original effective electronegativities $\chi_{\mathrm{eff},i}$. As a result, the external electric field $\boldsymbol{E}_{\textrm{ext,}i}$, depolarization field $\boldsymbol{E}_{\textrm{dep,}i}$, and their superposition $\boldsymbol{E}_{i}$ can be measured within molecular dynamics simulations by a repeated force measuring time step,
\begin{align}
&\boldsymbol{E}_{\textrm{ext,}i}  =  \frac{\boldsymbol{F}_{i}|_{\chi_{\textrm{eff,}i}}-\boldsymbol{F}_{i}|_{-\chi_{\textrm{eff,}i}}}{2q_{\textrm{pol,}i}},\label{eq:Eext Diagnostic}\\
&\boldsymbol{E}_{\textrm{dep,}i} = \frac{\boldsymbol{F}_{i}|_{\chi_{\textrm{eff,}i}}+\boldsymbol{F}_{i}|_{-\chi_{\textrm{eff,}i}}-2\boldsymbol{F}_{0,i}}{2q_{\textrm{pol,}i}},\label{eq:EDep Diagnostic}\\
&\boldsymbol{E}_{i} = \frac{\boldsymbol{F}_{i}|_{\chi_{\textrm{eff,}i}}-\boldsymbol{F}_{0,i}}{q_{\textrm{pol,}i}}.\label{eq:el.Feld Messung-1}
\end{align}

\subsection{Results and Discussion}

\begin{figure}[t!]
\begin{centering}
\includegraphics[width=8cm]{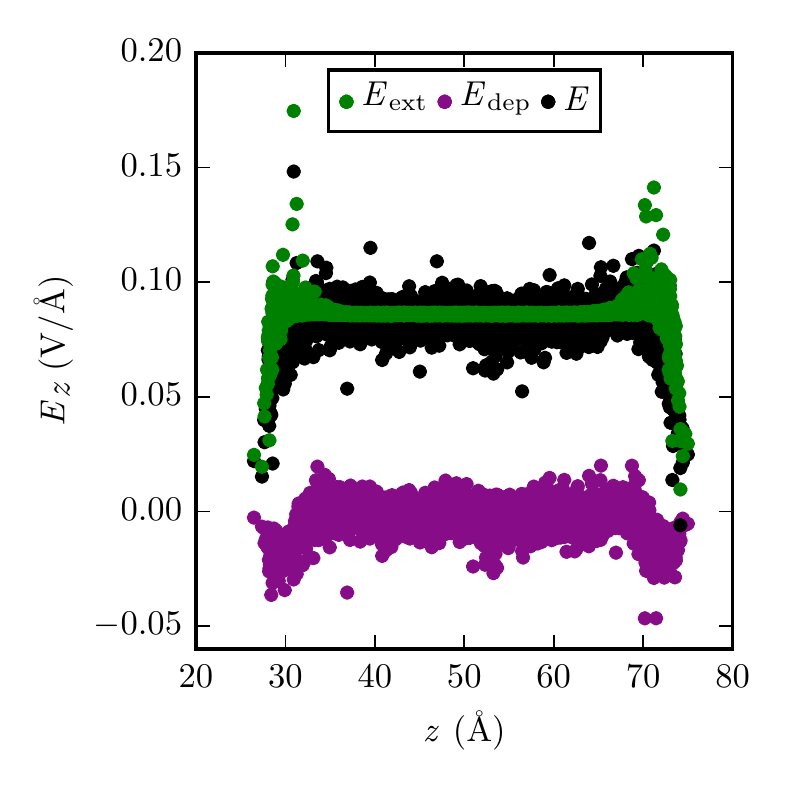}
\end{centering}
\caption{%
$z$ component of the external electric field, depolarization field, and the superposition of those at the atoms sites between the electrodes. \label{fig: QEq und QTPIE Felder}%
}
\end{figure}

This section covers the results concerning validation and consistency of the proposed model for an arbitrarily chosen voltage of 4~V. Fig. \ref{fig: QEq und QTPIE Felder} shows the projection of the $z$ component of the external electric field, depolarization field, and the superposition of both within the dielectric along the $z$ axis at the atom sites. The mean value of the external electric field within the dielectric fluctuates slightly around 0.08~V/\r{A}, which is in agreement with approximate calculations for the electric field strength of a parallel plate capacitor. In the vicinity of the Cu electrodes the electric field follows the negative gradient of the potential and is therefore decreased until it completely vanishes inside the electrodes. 

\begin{figure}[t!]
\begin{centering}
\includegraphics[width=8.0cm]{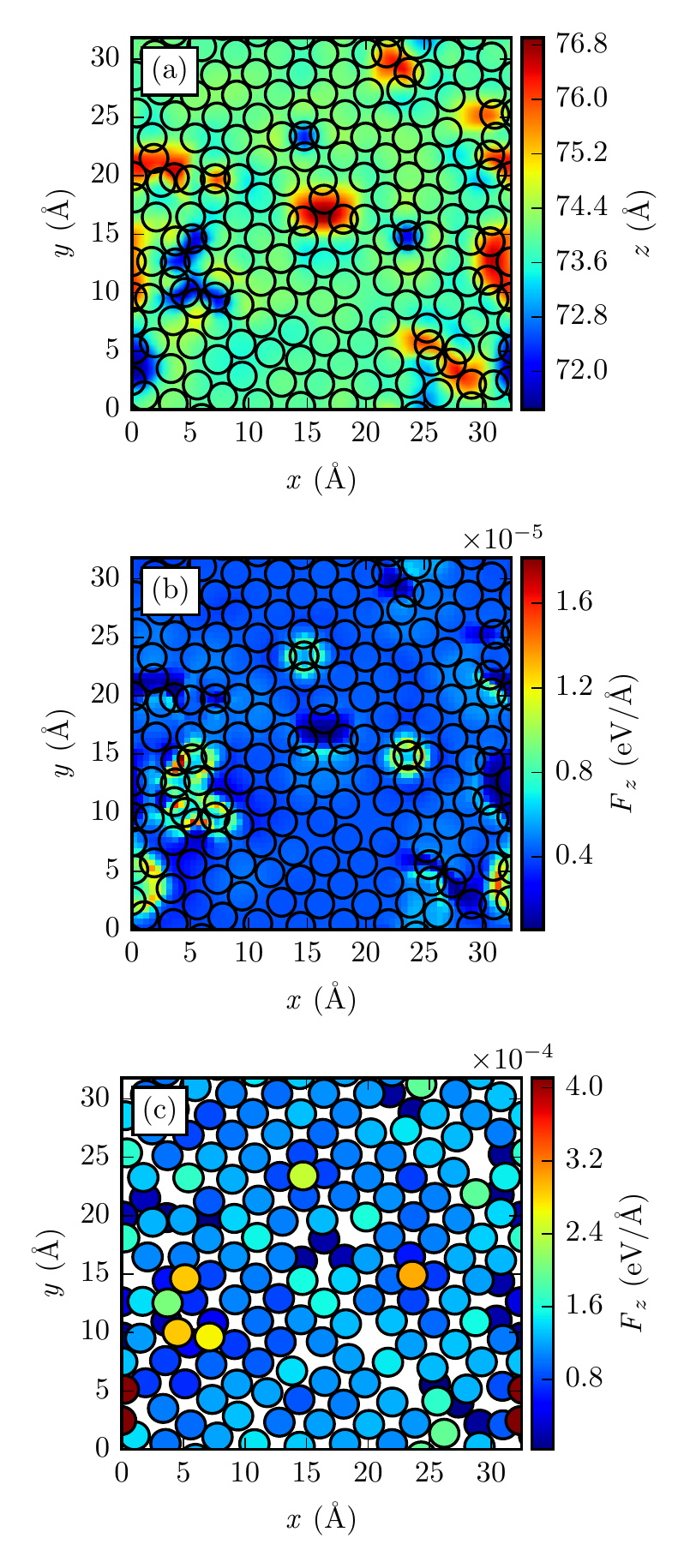}
\end{centering}
\caption{{(a) Interpolation of the metal atom sites at the interface, illustrated as circles. (b) Force in the $z$ direction on the free charges at the respective surface element due to the external electric field. (c) Resultant force in the $z$ direction per atom.} \label{fig: F an interfaces}}
\end{figure}

The calculated depolarization field fluctuates around 0~V/\r{A} within the a-SiO$_2$ due to a local compensation. The negative field component in the vicinity of the electrodes originates from the net contribution of the polarization charges. Due to the case of constant potentials a corresponding field and net charge is expected at the electrodes.  {The minor fluctuations within the a-SiO$_2$ are caused by interactions on the atomic scale and originate from the inherent restrictions of atomic point charge models, i.e., the lack of the spatially distributed charges per atom. The discrete charge distribution does consequently only allow for polarization along interatomic bonds, whereas any perpendicular contribution cannot be described and is therefore neglected \cite{chen2009theory}. The amorphous structure results in an equally disordered distribution of local polarization currents. The utilization of point charges leads thus to small fluctuations of the depolarization field in the $x-y$ plane. Overall, the resulting electric field within the system matches the expectations well.}
 At last the implementation of the forces on the free charges at the copper electrodes is verified.{ Fig. \ref{fig: F an interfaces} (a) shows the interfacial metal atoms which are bounded to the top electrode indicated as circles and the interpolation of their $z$ coordinates. Fig. \ref{fig: F an interfaces} (b) shows the force distribution on the free charges at the respective surface element due to the external electric field. A consistent relation between the interface gradient and the force distribution is expected due to the utilization of the former as Dirichlet boundary condition in the computation of the latter. The forces on the interfacial surface elements were subsequently assigned to the respectively nearest interfacial metal atom. The resulting force distribution per atom is shown in  Fig. \ref{fig: F an interfaces} (c). The overlap of atoms indicates their $z$ coordinate with respect to surrounding atoms. Atoms in the foreground have a lower relative $z$ coordinate. More surface elements are assigned to atoms in the foreground than to atoms in the background. The forces which are finally applied to the atoms are therefore a function of the respective overlap as well as interface gradient.} The closer the interfacial atom site towards the opposite electrode the greater the received force. The opposite can be observed for the inverse case.

The calculated electric field and force distribution in case of the chosen example system Cu/a-SiO$_2$/Cu are in agreement with fundamental field theoretical expectations. {This indicates that the mentioned assumptions within the model and the described approximations for the diagnostics of the electric fields are correct.}

For instance the influence of atomic rearrangement is neglected within the measurement of the depolarization field. However, the validation shows that those effects are of minor importance in this case. Thus, for material systems where the contribution of those mechanisms to the polarization is significant the presented diagnostics should be used with care.

\section{CONCLUSIONS\label{sec:CONCLUSIONS}}

A model of externally applied electric fields for molecular dynamics simulations and its fundamental implications to the atomic system for generic metal-insulator-metal systems has been proposed. {The model remedies certain drawbacks of existing models, e.g., restrictions to conductive dielectrics, non-reactive systems and thicknesses of the insulating material in the order of 10 \r{A}.} The extension of applicable systems originates from the force separation into long range electric forces and local polarization effects. As a consequence limitations due to the potential cutoff radius are bypassed. 

The model is applied to a resistive switching device, often realized as metal-insulator-metal system. It has been shown that the calculated electric field and force distribution in case of the chosen example system Cu/a-SiO$_2$/Cu are in agreement with fundamental field theoretical expectations. 

Future work on the basis of the proposed model for electric fields is planned. In particular, from a technical point of view an efficient and full implementation in LAMMPS is needed. From the physics point of view, we believe that an approximation for the Gaussian-type orbitals (GTOs) within the computation of the overlap integrals for the effective electronegativities should be realized, since GTOs are not needed for the computation of the respective Coulomb interactions.

\section*{ACKNOWLEDGMENTS\label{sec:ACKNOWLEDGMENTS}}

Financial support provided by the German Research Foundation DFG in the frame FOR 2093 ``Memristive Devices for Neuronal Systems'' and TRR 87 ``Pulsed High-Power Plasmas for the Synthesis of Nanostructured Functional Layers'' is gratefully acknowledged. Furthermore, T.G. thanks Frederik Schmidt for valuable discussions.

\section*{ORCID IDs}
\noindent\href{http://orcid.org/0000-0001-6445-4990}{T. Mussenbrock: http://orcid.org/0000-0001-6445-4990}

\newpage

%


\end{document}